\documentclass[aps,prc,preprint,groupedaddress]{revtex4}
\usepackage{amsmath} 
\usepackage{epsfig} 
\usepackage{graphics}
\begin{document}

\preprint{LA-UR--04-4790}

\title{Applying the Bloch-Horowitz equation to s- and p-shell nuclei}

\author{T.C.Luu}
\email[]{tluu@lanl.gov}
\affiliation{Los Alamos National Laboratory, MS-227, Los Alamos, New
  Mexico 87545, USA}

\author{P.Navratil} 
\email[]{navratil1@llnl.gov} 
\affiliation{Lawrence Livermore National Laboratory, L-414, P.O.Box
  808, Livermore, California 94551, USA}

\author{A.Nogga}
\email[]{nogga@phys.washington.edu}
\affiliation{Institute for Nuclear Theory, University of Washington,
  Box 351550, Seattle, Washington 98195, USA}

\date{\today}

\begin{abstract}
  The Bloch-Horowitz (BH) equation has been successfully applied to
  calculating the binding energies of the deuteron and $^3$H/$^3$He
  systems.  For the three-body systems, BH was found to be
  perturbative for certain choices of the harmonic oscillator (HO)
  parameter $b$. We extend upon this work by applying this formalism
  to the alpha particle and certain five-, six-, and seven-body nuclei
  in the p-shell.  Furthermore, we use only the leading order BH term
  and work in the smallest allowed included-spaces for each few-body
  system (0$\hbar\omega$ and 2$\hbar\omega$).  We show how to
  calculate A-body matrix elements within this formalism.  Stationary
  solutions are found for all nuclei investigated within this work.
  The calculated binding energy of the alpha particle differs by $\sim
  1$ MeV from Faddeev-Yakubovsky calculations.  However, calculated
  energies of p-shell nuclei are underbound, leaving p-shell nuclei
  that are susceptible to cluster breakup.  Furthermore, convergence
  is suspect when the size of the included-space is increased.  We
  attribute this undesirable behavior to lack of a sufficiently
  binding mean-field.
\end{abstract}

%\keywords{}

\maketitle

\section{Introduction\label{intro}}
Difficulties intrinsic to the many-body problem stem from the
exponentially-increasing complexity of the Hilbert space as the size
of the few-body system increases. This increasing complexity is very
daunting\footnote{A succinct enumeration of the complexity of the
few-body system is given in Table 2 of Ref.\cite{Pieper:2001mp}.}, yet
much progress has been made in developing methods that illuminate the
nuclear structure of few-body systems.  A particular method that
warrants mentioning is the Green's function Monte Carlo (GFMC) method
(see, for instance,
Refs.\cite{Carlson:1998qn,Pudliner:1997ck,Pieper:2001mp} and
references within).  Using bare `realistic' phenomenological two- and
three-body forces, GFMC calculations for certain p-shell nuclei have
produced nuclear binding energies to a precision better than 2\%.
This precision, coupled with the GFMC's success in reproducing
observed nuclear few-body data, has placed the GFMC method as the
benchmark which other methods use as comparison.  Yet even with the
consistent increases in computer resources and speed, GFMC
calculations are very computationally taxing.  GFMC calculations on
sd-shell nuclei, if possible, are still many years away.  Other
`infinite Hilbert-space' methods succumb to the same problems.

These difficulties have motivated the use of methods that employ
effective interactions within severely truncated Hilbert spaces, or
included-spaces, thereby alleviating many of the complexities of the
many-body system.  A classic example is the traditional nuclear shell
model (TNSM).  Unfortunately, calculating the effective interaction
exactly is as difficult a problem as the original `infinite
Hilbert-space' problem.  Hence some form of approximation must be done
to the effective interaction.  For example, in TNSM calculations the
effective interaction is approximated only at the two-body level (the
exact effective interaction is an A-body interaction, where A is the
number of nucleons). Its matrix elements are typically fitted to
existing nuclear data using a
G-matrix\cite{Bruckner:1955a,Bruckner:1955b} as a starting point.  An
alternative example (aiming at a convergence to the exact solution) is
the No-Core shell model
(NCSM)\cite{Navratil:1999pw,Navratil:2003ef,Navratil:2003ib}. In NCSM
\emph{ab initio} calculations, the effective interaction is derived
from the bare nucleon-nucleon (NN) interaction by performing a unitary
decoupling transformation.  NCSM calculations approximate the exact
A-body effective interaction by retaining only the two- or two- plus
three-body effective terms. Both methods have had resounding success
in describing nuclear few-body systems.  TNSM by and large remains the
most popular method for working with nuclei in the sd-shell, whereas
NCSM results for p-shell nuclei agree very well with GFMC.

In a recent paper by two of us, we applied an alternative effective
interaction given by the Bloch-Horowitz (BH) equation\cite{bloch:1958}
to the nuclear three-body system\cite{Luu:2004xc}.  We found that by
performing a certain ordering of diagrams, the BH equation could be
applied perturbatively within a finite range of the harmonic
oscillator (HO) parameter $b$.  In this paper, we extend upon this
work by applying the same formalism to the alpha particle and certain
p-shell nuclei.  Unfortunately, the next-to-leading-order and
higher-order terms in the BH expansion are increasingly difficult in
p-shell nuclei due to the various many-particle permutation operators
that come into play.  Hence we apply only the leading order BH term to
these nuclei.  Also, we work within the smallest allowed
included-spaces so as to alleviate as much as possible the complexity
of the few-body Hilbert spaces.

As mentioned in Ref.~\cite{Luu:2004xc}, the leading order BH term
resembles in some ways that of the G-matrix effective
interaction\cite{Bruckner:1955a,Bruckner:1955b}.  Yet there are
subtleties that differentiate the two expressions.  In this paper, we
further elaborate upon these differences, as well as make comparisons
to the `multi-valued' G-matrix of Ref.~\cite{Zheng:1995td}.  We show
that the leading order BH term is a natural extension of the G-matrix
in which spectator dependence and cluster recoil is explicitly taken
into account.  These new physical inputs are the main motivation for
using the leading-order BH term over the G-matrix.  Yet their
inclusion presents its own difficulties.  In particular, the leading
order BH expression is an A-body interaction.

In Sec.~\ref{BH} we give a cursory review of the Bloch-Horowitz
equation.  We describe our methodology and motivate our approximation
of the BH equation to its leading-order expression in
Sec.~\ref{approx. bh}.  In doing so, we compare the BH leading term to
the traditional G-matrix as well as the `multi-valued' G-matrix.
Since our BH term represents an A-body interaction, Sec.~\ref{a-body
matrix elements} describes how to calculate A-body matrix elements
using this term.  In particular, we show that the degrees of freedom
over spectator nucleons can be integrated out completely, leaving an
average residual two-body interaction.  Section~\ref{results} presents
our results for the alpha particle (as well as the three-body system)
and certain p-shell nuclei.  There are issues with regards to
convergence of our solutions when we increase our included-space
sizes.  We discuss these issues in Sec.~\ref{convergence} and compare
differences between our calculations and NCSM calculations.  We give
concluding remarks in Sec.~\ref{conclusion}.  To keep the presentation
concise, we reserve formal derivations for the appendices.

\section{Bloch-Horowitz equation\label{BH}}
The Bloch-Horowitz equation is 
\begin{equation}\label{eqn:bh1}
    H_{eff}(E)=P\left(H+H\frac{1}{E-QH}QH\right)P,
\end{equation}
where
\begin{equation}\label{eqn:h}
    H=\sum_{i<j}^A(T_{ij}+V_{ij})=\sum_i^AT_i+\sum_{i<j}^AV_{ij}
    -\frac{P^2_{c.m.}}{2M_A}.
\end{equation}
Here, the intrinsic Hamiltonian $H$ is obtained by summing the
relative kinetic and potential energy operators $T_{ij}$ and $V_{ij}$,
respectively, over all nucleon pairs $i$ and $j$.  The relative
kinetic energy is found by subtracting the center-of-mass (c.m.)
kinetic energy from the total single particle kinetic energies.  $M_A$
is the total mass of the A-body system.

In Eq.~\ref{eqn:bh1}, the operators $P$ and $Q=1-P$ are projection
operators that define the included- and excluded-spaces, respectively.
They are intimately related to the choice of basis states in which
calculations are performed.  For this work we chose the Jacobi HO
basis.  This basis, defined solely within the relative coordinates,
avoids the overcompleteness characterized by the the c.m.  motion
intrinsic to traditional slater-determinant HO bases.  By avoiding
this c.m.  motion, our original A-body problem becomes an (A-1)-body
problem.  The Jacobi included-space ($P$) in which we perform our
calculations is thus defined in terms of all relative-coordinate
configurations with HO energy quanta
$N\hbar\omega\leq\Lambda\hbar\omega$.  Hence $\Lambda$ represents the
demarcation between included- and excluded-spaces.

Though many of the complexities of the few-body problem are alleviated
by working in small included-spaces in which the BH equation is
defined, calculating matrix elements of $H_{eff}$ is still a very
daunting task, if not impossible.  The energy $E$ appearing in
Eq.~\ref{eqn:bh1}, for example, represents the exact eigenvalue of the
system.  Obviously, this eigenvalue is not known \emph{a priori},
hence the BH equation must be solved self-consistently.  Furthermore,
$H_{eff}$ represents an A-body interaction, regardless of the fact
that the original bare $H$ may have been an (n$<$A)-body interaction.
The most difficult aspect of the BH equation deals with the resolvent
appearing in Eq.~\ref{eqn:bh1}: $[1/(E-QH)]Q$.  This resolvent
represents the full many-body \emph{interacting} Green's function.
The $Q$ operators appearing in this expression implies that this
propagator resides in the excluded-space, which is of infinite size.
Clearly solving for this operator is as difficult as the original
`infinite-Hilbert space' few-body problem.  Hence some form of
approximation of the BH equation is in order.  The next section shows
how we appproximate the BH equation to a more tractable ladder
summation.

\section{Approximating the Bloch-Horowitz Equation\label{approx. bh}}
In Refs.\cite{Luu:2004xc,Haxton:2002kb} we presented arguments for
using a particular rearrangement of Eq.~\ref{eqn:bh1},
\begin{equation}\label{eqn:BHre}
H_{eff}=P\left\{\frac{E}{E-TQ}[T_{eff}+V_{eff}]\frac{E}{E-QT}\right\}P,
\end{equation}
where
\begin{eqnarray}
T_{eff} &=& T+T\frac{-1}{E}QT, \label{eqn:Teff}\\ V_{eff} &=&
V+V\frac{1}{E-QH}QV.\label{eqn:Veff}
\end{eqnarray}
We stress that this particular form of the effective interaction comes
from a formal manipulation of the BH equation.  At this stage no
approximations have been introduced.  Equation~\ref{eqn:BHre} is
convenient to work with since it separates out an effective kinetic
term and an effective potential term.  As argued in
Refs.\cite{Luu:2004xc,Haxton:2002kb}, the operator $[E/(E-QT)]P$ that
sandwiches $T_{eff}$ and $V_{eff}$ can be used to define new basis
states that have the correct exponential asymptotic behavior, as
opposed to the Gaussian decay of HO wavefunctions.  This corrects for
HO `overbinding'.  Furthermore, matrix elements of $T_{eff}$ are
relatively easy to calculate.  In Ref.\cite{Luu:2004xc} we gave the
analytic expression for calculating three-body matrix elements of
$T_{eff}$.  In Appendix~\ref{Teff abody} we generalize this expression
to A-body systems.

The crux of the problem deals with $V_{eff}$, since the same
troublesome resolvent described in the previous section appears in
this operator.  For the three-body system of Ref.\cite{Luu:2004xc}, we
invoked Faddeev decompositions to generate a perturbative expansion of
the $V_{eff}$.  In what follows, we show an alternative approximation
to $V_{eff}$ which is completely equivalent to the leading-order term
of the expansion of Ref.\cite{Luu:2004xc}, but generalized to A-body
systems.

\subsection{Retaining the ladder summation\label{ladder sum}}
\begin{figure}
  \epsfig{file=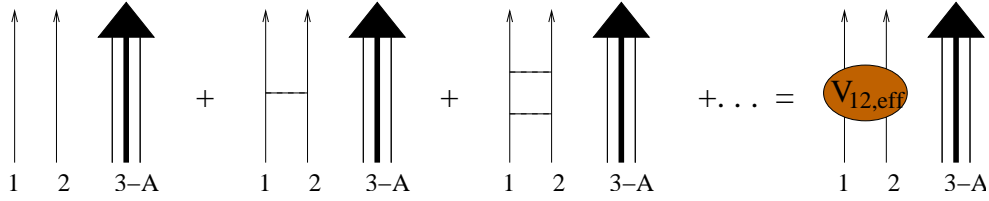,width=13cm}
  \caption{(Color online) Diagrammatic terms kept in ladder
    approximation of Eq.~\ref{eqn:G expansion}. Nucleons 1 and 2 are
    interacting via a ladder sum in the presence of spectator nucleons
    3-A.  Dashed lines represent interactions with
    $V_{12}$. \label{fig1}}
\end{figure}

We begin by expanding the resolvent in powers of $QV$,
\begin{equation}\label{eqn:G expansion}
  \frac{1}{E-QH}Q=\frac{1}{E-QT}Q+\frac{1}{E-QT}QV\frac{1}{E-QT}Q+
  \frac{1}{E-QT}QV\frac{1}{E-QT}QV\frac{1}{E-QT}Q+\ldots
\end{equation}
We retain from this expansion only the terms that contribute to a
ladder summation, as shown diagrammatically in Fig.~\ref{fig1}.  Here
only particles 1 and 2 are interacting via repeated insertions of the
potential $V_{12}$, while particles 3 through A are spectator nucleons.
This infinite ladder sum is collectively grouped into the expression
$V_{12,eff}$.  Hence $V_{eff}$ is approximated as
\begin{eqnarray}\label{eqn:approxVeff}
        V_{eff}&\sim&\sum_{i<j}^AV_{ij,eff}\\
             &=&\frac{A(A-1)}{2}V_{12,eff}, \label{eqn:approxVeff2}
\end{eqnarray}
where the $A(A-1)/2$ factor comes from using anti-symmetrised states.
Obviously the choice of $ij=12$ in Eq.~\ref{eqn:approxVeff2} is
arbitrary.

There are three distinct ways in which to do the ladder summation.
The first method assumes that the spectator particles are completely
irrelevant.  They do not contribute in any way to the ladder
interactions between particles 1 and 2.  Such an approximation is akin
to the G-matrix interaction of traditional shell-model calculations.
The resulting interaction is completely two-body in nature which, from
a computational standpoint, greatly simplifies the problem.
Furthermore, the effective interaction is much softer than the
original bare $NN$ interaction, thereby reducing the coupling of the
high momentum matrix elements to the included space. This softening of
the core is a generic property of the ladder sum.  Indeed, it can be
shown that in the limit of an infinite hard core, matrix elements of
this ladder summation are still finite\cite{Fetter&Walecka:1971}.  With these
approximations, Eq.~\ref{eqn:G expansion} becomes
\begin{multline}\label{eqn:regG}
  V_{12,eff}=\frac{1}{E-Q_{12}T_{12}}Q_{12}+
  \frac{1}{E-Q_{12}T_{12}}Q_{12}V_{12}\frac{1}{E-Q_{12}T_{12}}Q_{12}+\\
  \frac{1}{E-Q_{12}T_{12}}Q_{12}V_{12}
  \frac{1}{E-Q_{12}T_{12}}Q_{12}V_{12}\frac{1}{E-Q_{12}T_{12}}Q_{12}+\ldots
\end{multline}
Note the replacement of $Q$ with $Q_{12}$ in the previous equation.
Since the presence of spectator nucleons is completely ignored, the
many-body projection operator $Q$ acts only within the two-body
sector.  This condition is valid only in the $\Lambda=0$ excitation
spaces\cite{Zheng:1995td}. We label this operator as $Q_{12}$.  Since
this interaction is only two-body, it is a poor approximation to the
original $H_{eff}$, since the latter is an A-body interaction.
Furthermore, since $T$ is replaced by $T_{12}$, the energy E can no
longer be associated with the energy of the A-body system.
Historically, the missing many-body correlations have been `mocked' up
by introducing fictitious parameters, such as starting energies
$\omega$.  These further approximations eliminate any
self-consistency, as well as manifest A-body anti-symmetry.  Attempts
at improving the effective interaction via perturbation have had mixed
success as well.  Barrett and Kirson\cite{Barrett:1970} have shown
that matrix elements of third order (in G) diagrams can be as large as
second order diagrams.  These inconsistencies have led shell model
practitioners on a more phenomenological track, using effective
two-body matrix elements derived by fitting to nuclear spectra.
Clearly any underlying connection to the original bare $NN$
interaction has been lost at this point.

Since the included space is defined in terms of many-body states (and
not just two-body states) with total HO quanta below $\Lambda$, the
actual HO configuration of the spectator nucleons restricts the
possible configuration in which the two-nucleon cluster may interact.
This is especially true for multi-valued excitation spaces
(\emph{i.e.} $\Lambda>0$).  For example, within a
$\Lambda\hbar\omega=4\hbar\omega$ included space, a $2\hbar\omega$
spectator state restricts the possible two-body interacting cluster to
either $0\hbar\omega$ or $2\hbar\omega$, since the sum of the HO
quanta of the spectator and interacting clusters must add to a value
$\le\Lambda=4$. The multiplicity of the two-nucleon state implies that
the G-matrix is multivalued.  This implicit spectator dependence can
be included within the ladder sum by replacing $Q_{12}$ by the full
many-body $Q$ within Eq.~\ref{eqn:regG},
\begin{multline}\label{eqn:multiG}
  V_{12,eff}=\frac{1}{E-QT_{12}}Q+
        \frac{1}{E-QT_{12}}QV_{12}\frac{1}{E-QT_{12}}Q+\\
                \frac{1}{E-QT_{12}}QV_{12}
                \frac{1}{E-QT_{12}}QV_{12}\frac{1}{E-QT_{12}}Q+\ldots
\end{multline}
Equation~\ref{eqn:multiG} represents the second method for performing
the ladder summation.  Zheng \emph{et al.}\cite{Zheng:1995td} have
shown that there are significant improvements to using the
multi-valued G-matrix over its single-valued cousin.  In particular,
the calculated low energy excitation spectrum of certain few-body
systems were lowered relative to the ground state when using the
multi-valued G, bringing energies into better agreement with
experiment.  Despite its spectator dependence, however,
Eq.~\ref{eqn:multiG} still represents a two-body operator since it is
diagonal in all spectator quantum numbers.  Hence induced
multi-particle correlations are still missing.

In the center of mass frame, the relative kinetic energies of the
interacting two-nucleon cluster and spectator cluster are correlated.
This is especially true for light systems.  This cluster recoil can be
incorporated into the ladder sum by using the full kinetic energy
operator $T$ in Eq.~\ref{eqn:multiG},
\begin{multline}\label{eqn:recoilG}
  V_{12,eff}=\frac{1}{E-QT}Q+
  \frac{1}{E-QT}QV_{12}\frac{1}{E-QT}Q+\\
  \frac{1}{E-QT}QV_{12}
  \frac{1}{E-QT}QV_{12}\frac{1}{E-QT}Q+\ldots
\end{multline}
Note that since the full $Q$ and $T$ are present,
Eq.~\ref{eqn:recoilG} represents a multi-valued, \emph{A-body}
operator.  The A-body nature of this operator implies that some
induced multi-particle correlations are being incorporated at this
level of approximation.  However, these correlations do not come from
potential interactions; rather, they stem solely from the fact that
the kinetic energies of the clusters are correlated.  This is the
third and final approximation to $V_{eff}$ (Eq.~\ref{eqn:Veff}) which
can be obtained from a ladder summation.  It is also the approximation
of $V_{eff}$ that we investigate in this paper.

\section{Calculating A-body matrix elements of $V_{12,eff}$
  \label{a-body matrix elements}} 
The infinite ladder sum in Eq.~\ref{eqn:recoilG} can be expressed in
closed form\cite{Luu:2004xc},
\begin{equation}\label{eqn:ladderClosedForm}
  V_{12,eff}=t_{12}-t_{12}G_0P\frac{1}{\Gamma_0+\Gamma_\infty}PG_0t_{12},
\end{equation}
where
\begin{eqnarray}
  G_0&=&\frac{1}{E-T},\\ \label{eqn:G0}
  \Gamma_0&=&PG_0P,\\ \label{eqn:Gamma0}
  t_{12}&=&V_{12}+V_{12}G_0t_{12},\\ \label{eqn:t12}
  \Gamma_\infty&=&PG_0t_{12}G_0P. \label{eqn:Gamma_inf}
\end{eqnarray}
This is precisely the leading-order BH term of Ref.\cite{Luu:2004xc}.
Here $G_0$ represents the free many-body propagator, while $\Gamma_0$
consists of included-space overlaps of $G_0$.  The t-matrix for free
particles satisfies the Lippman-Schwinger equation that depends on
$V_{12}$ and the many-body propagator $G_0$, and $\Gamma_\infty$
represents included-space overlaps of $t_{12}$ weighted by two
insertions of $G_0$.

The many-body nature of the operator $V_{12,eff}$ comes directly from
the included-space overlaps of $G_0t_{12}G_0$.  In momentum space
$t_{12}$ is not only a function of the interacting two-nucleon
cluster's initial and final momenta, $p$ and $p'$ respectively, but
also of the individual spectator Jacobi momenta $q_i$,
\begin{equation}\label{momentum t12}
  t_{12}(p',p; E-\sum_{i=1}^{A-2}q_i^2/2M_n).
\end{equation}
Here $M_n$ is the nucleon mass. This dependence on spectator momenta
implies that overlaps of this operator are not diagonal in all the
spectator's quantum numbers, and in particular, their principle
quantum numbers can change by arbitrary amounts.  In principle,
calculating included-space many-body matrix elements of this operator
would require numerically evaluating A-dimensional nested integrals in
momentum space (the same is true in coordinate space).  Such integrals
can be done via Monte Carlo methods, though these methods would be
very computationally taxing and probably inaccurate due to the
oscillatory nature of the HO wavefunctions.  In Appendix~\ref{int
spect} we show an alternative method in which the multi-dimensional
integrals over the spectator nucleons can be reduced to a sum over
one-dimensional integrals.  The result is valid regardless of the
number of spectator nucleons, and is given by
\begin{multline}\label{eqn:abodyt12}
  \int dq_1\cdots dq_{A-2}\ q_1^2\cdots q_{A-2}^2\ 
  R_{n_1'l_1}(q_1)R_{n_1l_1}(q_1)\cdots
  R_{n_{A-2}'l_{A-2}}(q_{A-2})R_{n_{A-2}l_{A-2}}(q_{A-2})\\
  \times t_{12}(p',p;E-\sum_{i=1}^{A-2} \frac{q_i^2}{2 \mu})=\\
  \sum_{m_1,m'_1,\cdots,m_{A-2},m'_{A-2}=0}^{n_1,n'_1,\cdots,n_{A-2},n'_{A-2}}
  C(m_1,m'_1,\cdots,m_{A-2},m'_{A-2})\\
  \times\left[\int_0^{\infty} d\rho\ \rho^{A-3+2\Sigma_i^{A-2} 
      (l_i+m_i+m_i'+1)}e^{-\rho^2}
    t_{12}(p',p;E-\frac{\rho^2}{2M_n b^2})\right],
\end{multline}
where $b$ is the oscillator parameter and
\begin{multline}\label{eqn:Ccoeff}
  C(m_1,m'_1,\cdots,m_{A-2},m'_{A-2})=\\
  (-1)^{\Sigma_i(n_i+m_i)}\left[\prod_{i=1}^{A-2}\
    \frac{\sqrt{2\Gamma(n_i+1)\Gamma(n_i+l_i+3/2))}}
    {\Gamma(n_i-m_i+1)\Gamma(l_i+m_i+3/2)\Gamma(m_i+1)}\right]\\
  \times (-1)^{\Sigma_i(n'_i+m'_i)}\left[\prod_{i=1}^{A-2}\
    \frac{\sqrt{2\Gamma(n'_i+1)\Gamma(n'_i+l_i+3/2))}}
    {\Gamma(n'_i-m'_i+1)\Gamma(l_i+m'_i+3/2)\Gamma(m'_i+1)}\right]\\
  \times\left[  \prod_{i=2}^{A-2}
    \frac{\Gamma(l_i+m_i+m_i'+3/2)\Gamma(\sum_{j=1}^{i-1}(l_j+m_j+m_j'+1)+
      \frac{i-1}{2})}
    {2\Gamma(l_i+m_i+m_i'+\sum_{j=1}^{i-1}(l_j+m_j+m_j'+1)+\frac{i+2}{2})}
    \right].
\end{multline}
In Eq.~\ref{eqn:Ccoeff} the first two products run over all spectator
nucleons (\emph{i.e.} $i$=1 to $A$-2).  Equation~\ref{eqn:abodyt12}
utilized the fact that matrix elements of $t_{12}$ are diagonal in the
spectator's \emph{orbital} quantum numbers $l_i$ (see
Appendix~\ref{int spect}), but not their principle quantum numbers.
The right hand side (\emph{RHS}) of Eq.~\ref{eqn:abodyt12} represents
a significant simplification to the original multi-dimensional
integrals, since the one-dimensional integrals of the \emph{RHS} can
be done quickly and accurately.  The coefficients
$C(m_1,m'_1,\cdots,m_{A-2},m'_{A-2})$ are known analytically and can
be pre-computed and stored.

\subsection{Dependence of $t_{12}$ on spectator nucleons\label{mf interaction}}
Equation~\ref{eqn:abodyt12} is only a function of the initial and
final momenta of the interacting two-nucleon cluster (\emph{i.e.}
nucleons one and two).  However, its functional form depends
implicitly on the number of spectator nucleons and their initial HO
momentum distribution.  In some sense, $\overline{t_{12}}$ represents
the average interaction of particles one and two under the influence
of the spectator's momentum distributions.  It is not a mean-field
interaction, since $\overline{t_{12}}$ is not derived from any
\emph{interactions} with the spectator nucleons.  We will discuss this
point in more detail later.

\begin{figure}
  \epsfig{file=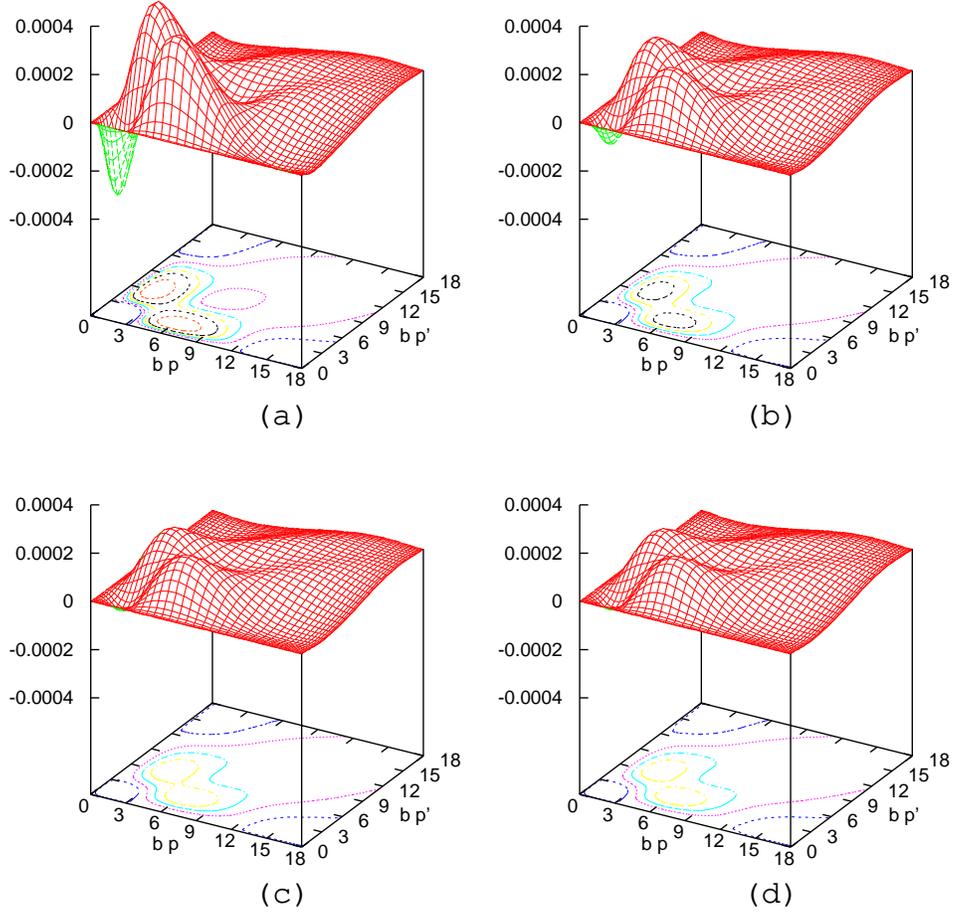,width=13cm}
  \caption{(Color online) Spectator dependence of
  $\overline{G_0t_{12}G_0}$.  All axes are dimensionless.  Plot (a) is
  calculated with no spectator nucleons.  Plot (b) includes two extra
  spectator nucleons on top of the interacting pair.  In this case,
  the Jacobi relative angular momentum for these nucleons are in the
  $0s$ states.  Plot (c) and (d) include a third and fourth nucleon,
  respectively, in the Jacobi relative $0p$ states.
    \label{fig2}}
\end{figure}

Figure~\ref{fig2} shows this spectator dependence of the operator
$b \hbar\omega p'^2\overline{G_0(p';E)t_{12}(p',p;E)G_0(p;E)}p^2$,
which is dimensionless.  Note that $\omega$ is the HO frequency,
defined by $\hbar\omega=\hbar^2/M_Nb^2$, where $M_N$ is the nucleon
mass. We plot this quantity rather than $\overline{t_{12}(p',p;E)}$
since it is more relevant for our calculations and more visually
instructive.  All plots within Fig.~\ref{fig2} are for the case
when $t_{12}$ is in the $^1S_0$ channel, plotted against initial and
final dimensionless momenta $bp$ and $bp'$, respectively.
Furthermore, all plots were calculated with E=-20 MeV and $b$=1.2 fm.
Plot (a) shows this operator when there are no spectator nucleons
present.  In this case $t_{12}$ is the standard two-body t-matrix.
Plot (b) shows how this operator is modified after integrating out the
presence of two spectator nucleon in the lowest allowed relative Jacobi
configuration, the $0s$ state.  Plot (c) shows the effect of
integrating out three spectator nucleons, two of which are in $0s$
states and one in a $0p$ state.  Finally, plot (d) shows how the
interaction is altered with the presence if four spectator nucleons
are integrated out.  In this case, two of the nucleons are in relative
$0s$ states, and two are in the relative $0p$ states.

The physical interpretations of the troughs and peaks seen in
Fig.~\ref{fig2} are difficult to explain due to the complicated
structure of the operator
$\overline{G_0(p';E)t_{12}(p',p;E)G_0(p;E)}$.  What is clear is that
the inclusion of additional spectator nucleons softens this operator.
This is consistent with the fact that the possible configurations
of the interacting two-nucleon cluster are reduced with increasing
numbers of spectator nucleons.  

\section{Results\label{results}}
In all our calculations we use the Argonne $v_{18}$ and $v'_8$
potentials\cite{Wiringa:1995wb} (Av18 and Av8') for our $NN$
interactions.  Furthermore, we ignore isospin symmetry breaking
effects included in the Av18 potential since this is a small effect.
Table~\ref{tab:dimensions} shows the dimensions of the $\Lambda=0,2$
included-spaces for the ground state quantum numbers of different
few-body systems considered in this paper.  Since the dimensions are
very small, our calculations do not demand extravagant computing
resources and can be done very quickly.  To construct our
included-space anti-symmetrised HO Jacobi wavefunctions, we use a
modified version of the routine \emph{manyeff} outlined in
Ref.\cite{Navratil:1999pw}.  We make our comparisons with GFMC and
NCSM calculations, and show experimental results solely as reference
points.

\subsection{S-shell nuclei\label{s-shell}}
\begin{figure}
\begin{center}
  \epsfig{file=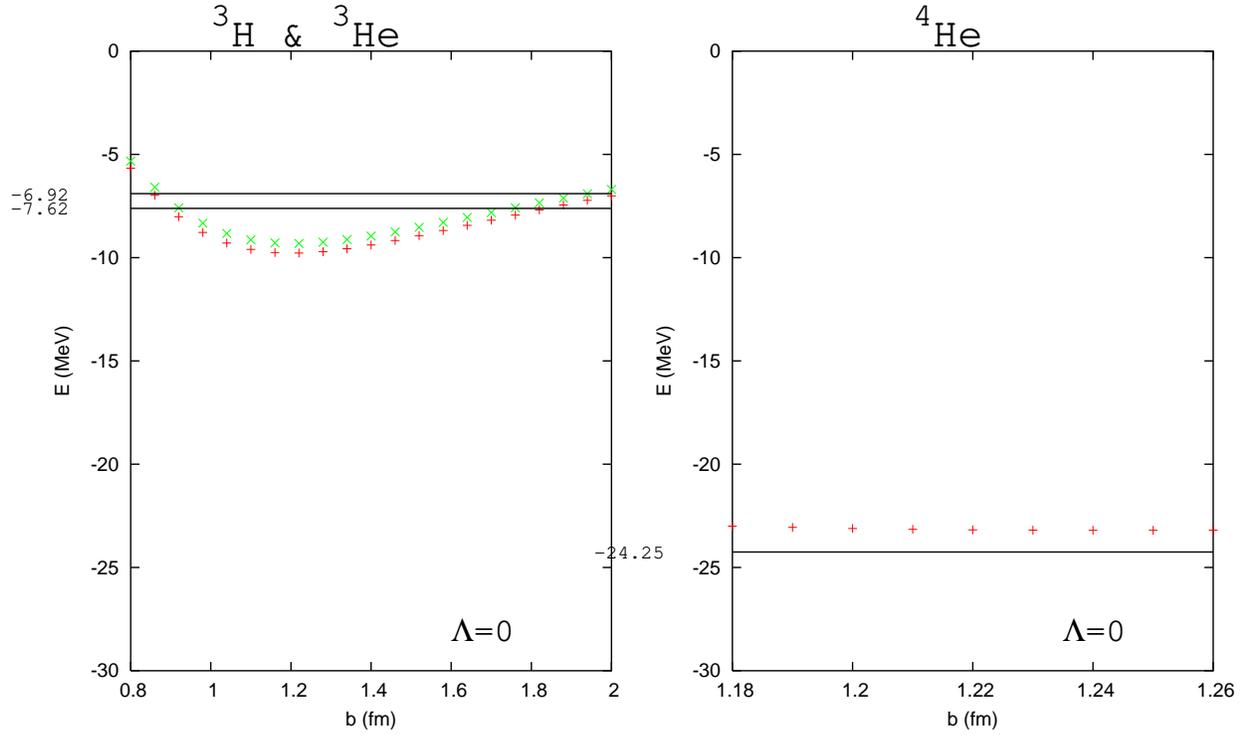,width=16cm}
  \caption{(Color online) S-shell results for $\Lambda=0$ included
    spaces.  Points correspond to our calculations, whereas solid
    lines correspond to the `exact' results taken from
    Ref.\cite{Nogga:2001cz}. In the three-body system, the upper and
    lower set of points refer to $^3$He and $^3$H, respectively.  We
    do not show results for $^2$H, since all calculated points lie on
    the exact theoretical result, regardless of $b$.  Calculations
    were done with the Av18 potential.\label{fig3}}
\end{center}
\end{figure}

Figure~\ref{fig3} shows our results of using $V_{12,eff}$ on s-shell
nuclei.  All calculations were done in $\Lambda=0$ included spaces and
with the Av18 potential.  Our results are displayed by points, while
the `exact' theoretical results are taken from Ref.\cite{Nogga:2001cz}
and displayed as solid lines.  For the three-body system, the upper
and lower set of points refer to $^3$He and $^3$H, respectively.  Note
that for $^4$He, the plotted range of $b$ values is much smaller.

The $^2$H system is special since in this two-body system there are no
spectator nucleons.  The ladder sum of Fig.~\ref{fig1} is exact in
this case, meaning that $V_{12,eff}$ is not an approximation but
corresponds to the \emph{exact} effective interaction.  Thus any
calculations of the two-body system cannot depend on $b$, as this is
an intrinsic parameter of the HO basis and has nothing to do with the
$NN$ interaction.  Indeed, all our $^2$H calculations lie on the exact
theoretical binding energy\footnote{The theoretical binding energy
  also corresponds to the experimental binding energy, since this is
  an input in the construction of the Av18 potential.}, regardless of
the oscillator parameter $b$ (as well as size $\Lambda$).  Hence, we
do not show this case in Fig.~\ref{fig3}.

The variation of the binding energy with $b$ in the three-body system
and higher is a key indicator that $V_{12,eff}$ is only approximative
for these systems.  Furthermore, the fact that the calculated binding
energies can overbind for certain values of $b$, as shown in the
three-body system, shows that this approximation is not variational.
For the alpha particle, our results underbind by $\sim 1$ MeV.

\subsection{P-shell nuclei\label{p-shell}}
\begin{figure}
  \epsfig{file=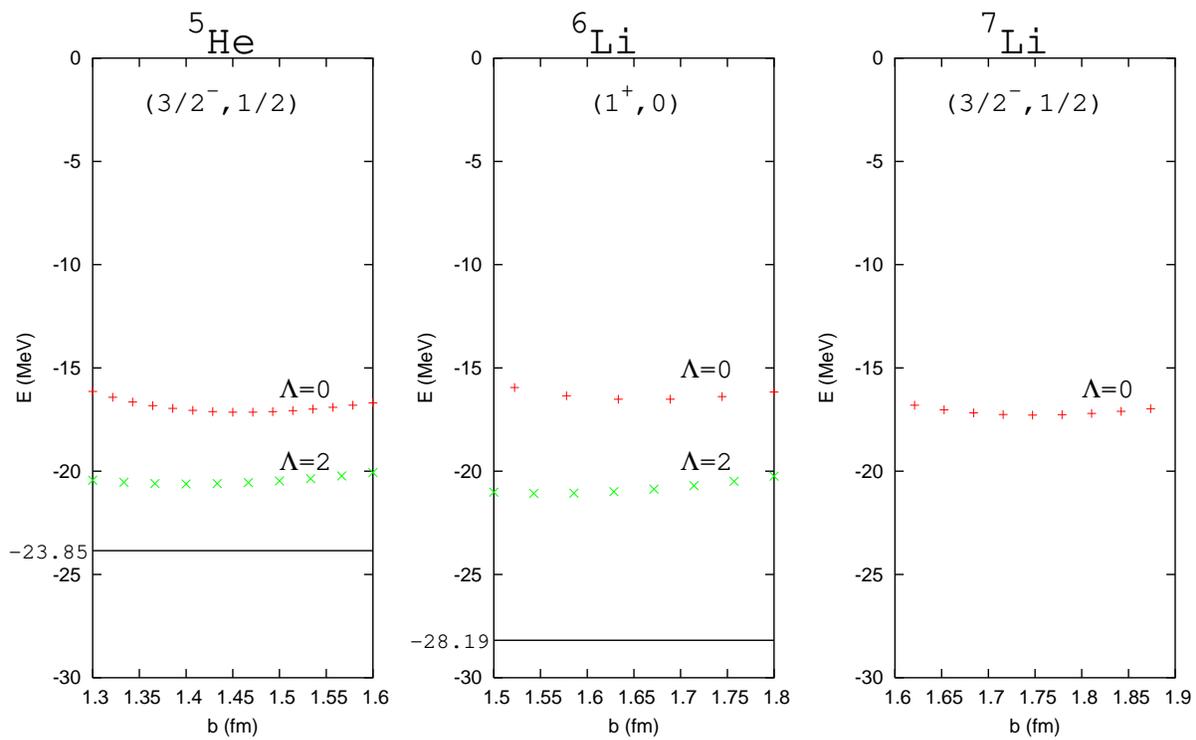,width=16cm}
  \caption{(Color online) Same as in Fig.~\ref{fig3}, but for p-shell
  ground states using Av8'. Solid lines correspond to results taken from
  Ref.\cite{Pieper:2004qw}.\label{fig4}}
\end{figure}

\begin{figure}
  \epsfig{file=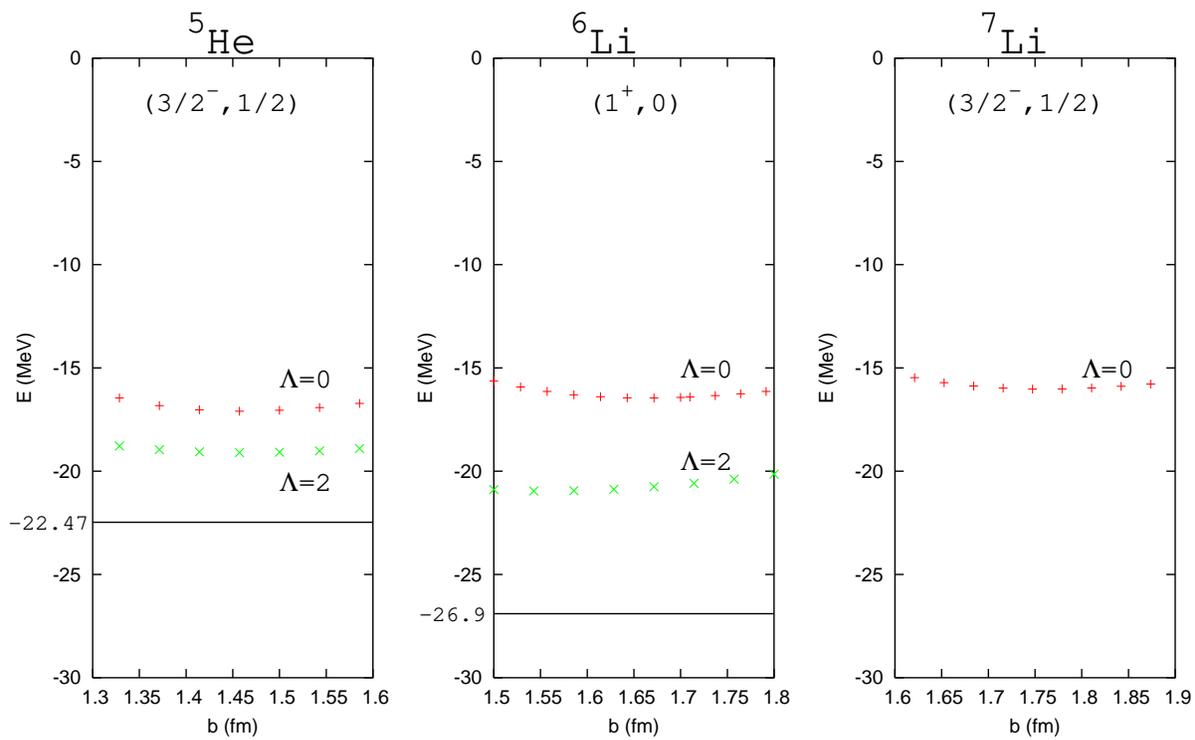,width=16cm}
  \caption{(Color online) Same as in Fig.~\ref{fig4}, but using Av18.
  \label{fig5}}
\end{figure}

Figure~\ref{fig4} shows our results for the ground state binding
energies for $^5$He, $^6$Li, and $^7$Li using the Av8'
potential\footnote{Our implementation of the Av8' potential includes
  the isoscalar Coulomb terms.}. Again, our results are displayed with
points, while the `exact' theoretical binding energies are taken from
Refs.\cite{Pieper:2004qw}  and displayed by solid lines.  For the seven-body
system, the calculated GFMC binding energy is -33.56 MeV, which is
below the plotted range in Fig.~\ref{fig4}.  Numbers in parenthesis
refer to the quantum numbers (J$^\pi$,T).  For the five- and six-body
systems, we show calculations for both $\Lambda=0$ and $\Lambda=2$
included spaces.  For the p-shell nuclei presented here our results
underbind for all values of $b$.  We plot our results where our
calculated binding energies approach closest to the exact theoretical
results.  The selected ranges of $b$ in Fig.~\ref{fig4} reflect this
fact.  Figure~\ref{fig5} shows the same results using the Av18
potential.

Since the $\Lambda=0$ calculations of these p-shell nuclei all
underbind, they are susceptible to cluster breakup.  Furthermore,
their calculated binding energies are nearly degenerate.  We do not
have a satisfactory explanation for this behavior.  This apparent
degeneracy is broken in the $\Lambda=2$ calculations of the five- and
six-body systems.  We have not performed any $\Lambda=2$ calculations
for the seven-body system due to time and computer limitations.

\begin{figure}
  \epsfig{file=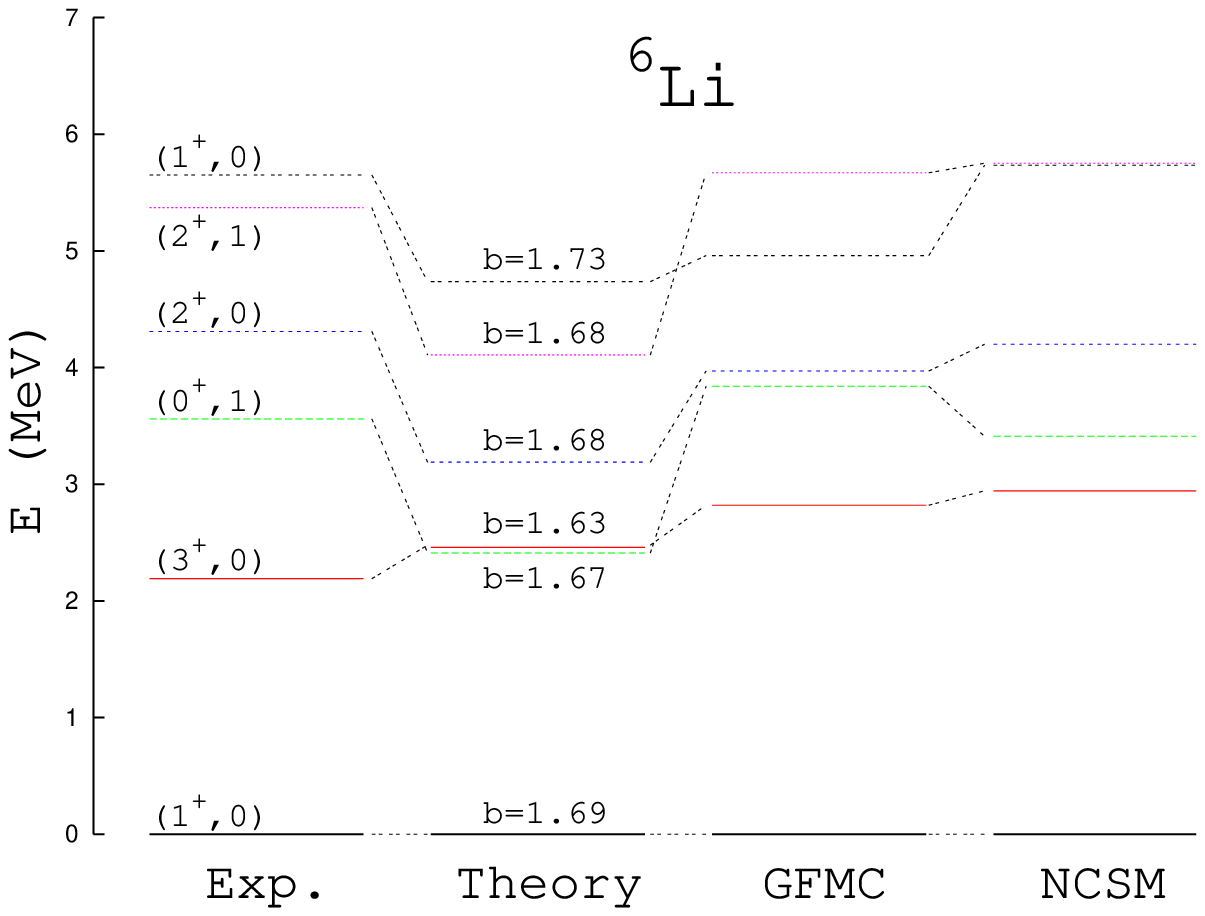,width=12cm}
  \caption{(Color online) Excited spectrum of $^6$Li using
    Av8'.\label{fig6}}
\end{figure}

\begin{figure}
  \epsfig{file=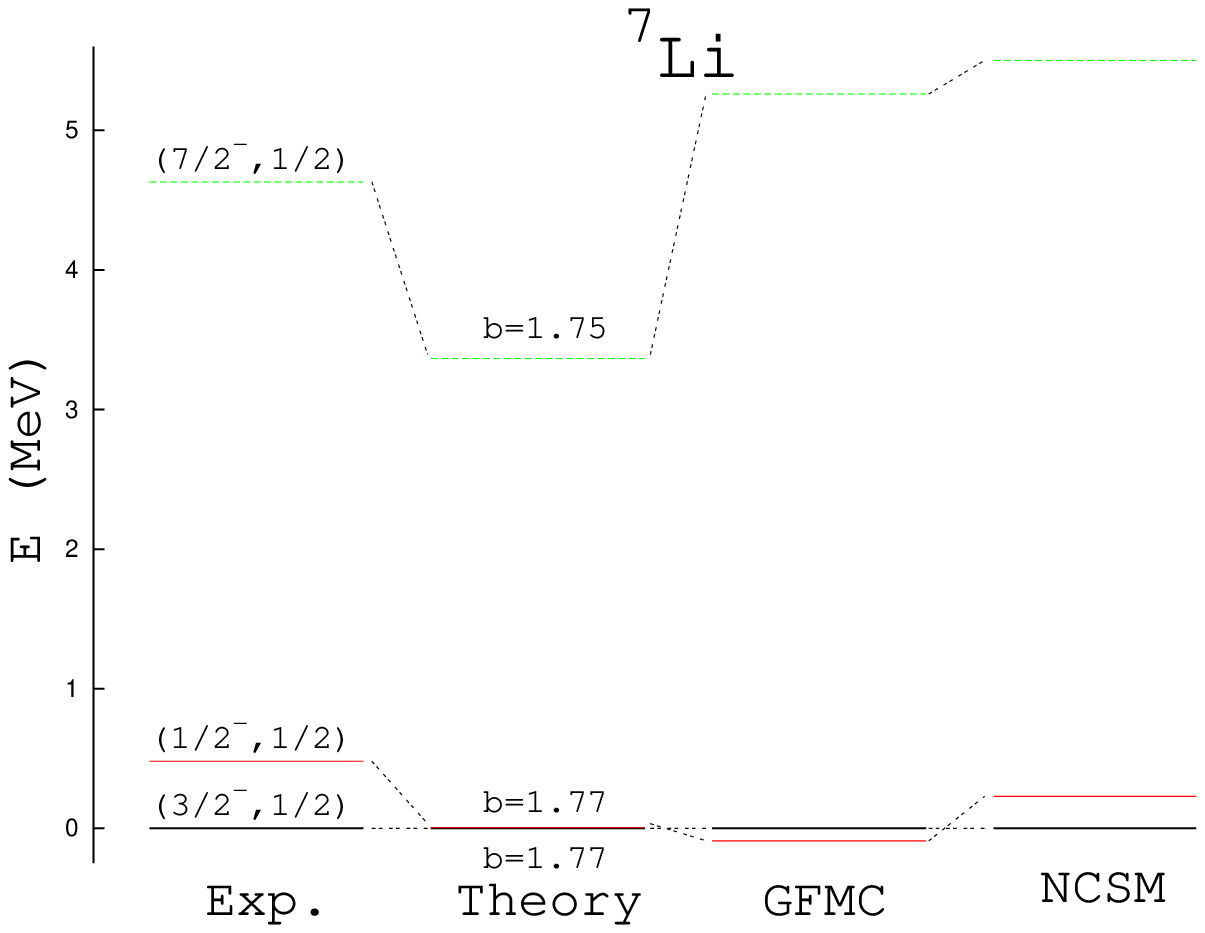,width=12cm}
  \caption{(Color online) Excited spectrum of $^7$Li using
    Av8'.\label{fig7}}
\end{figure}

\begin{figure}
  \epsfig{file=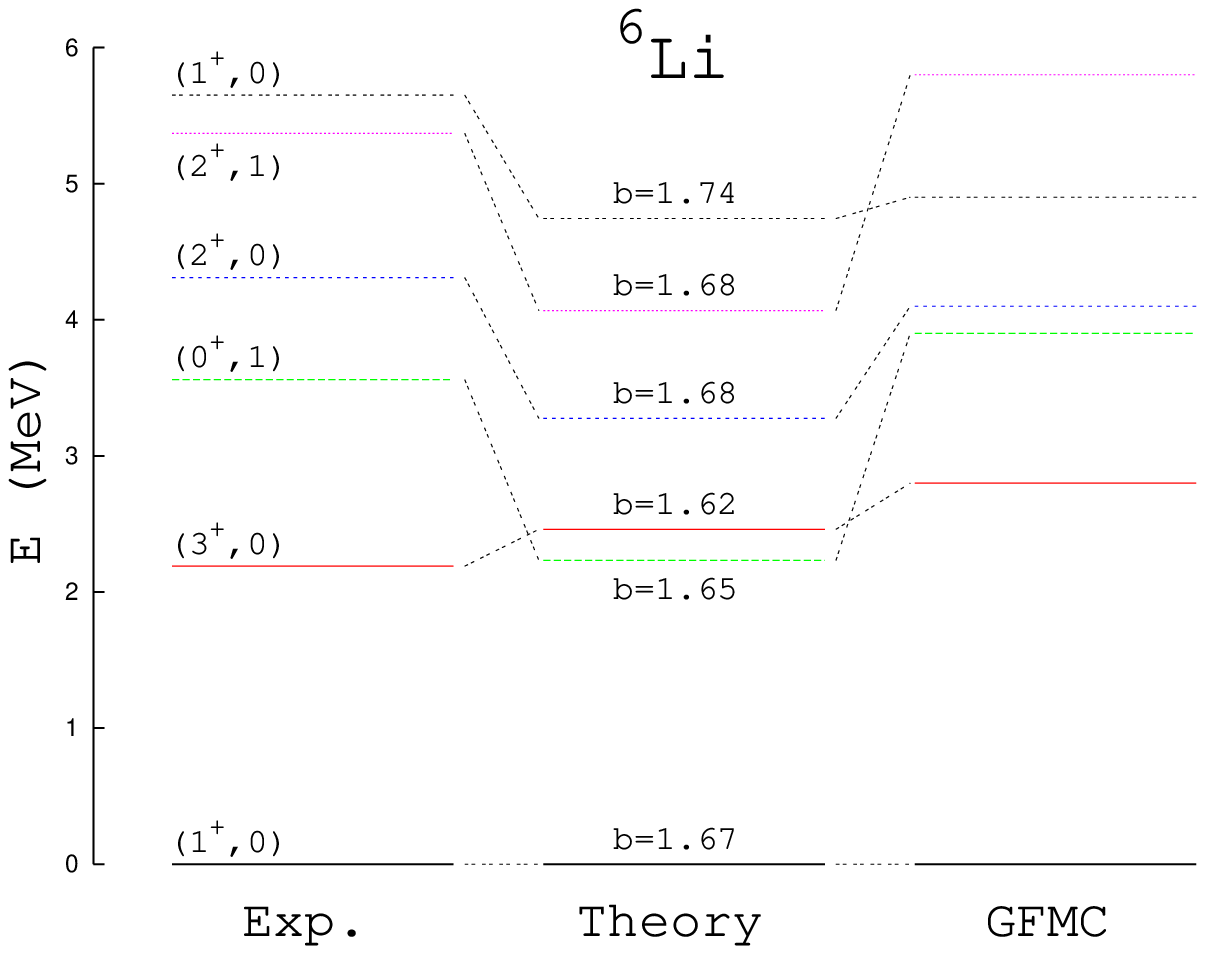,width=12cm}
  \caption{(Color online) Excited spectrum of $^6$Li using
    Av18.\label{fig8}}
\end{figure}

\begin{figure}
  \epsfig{file=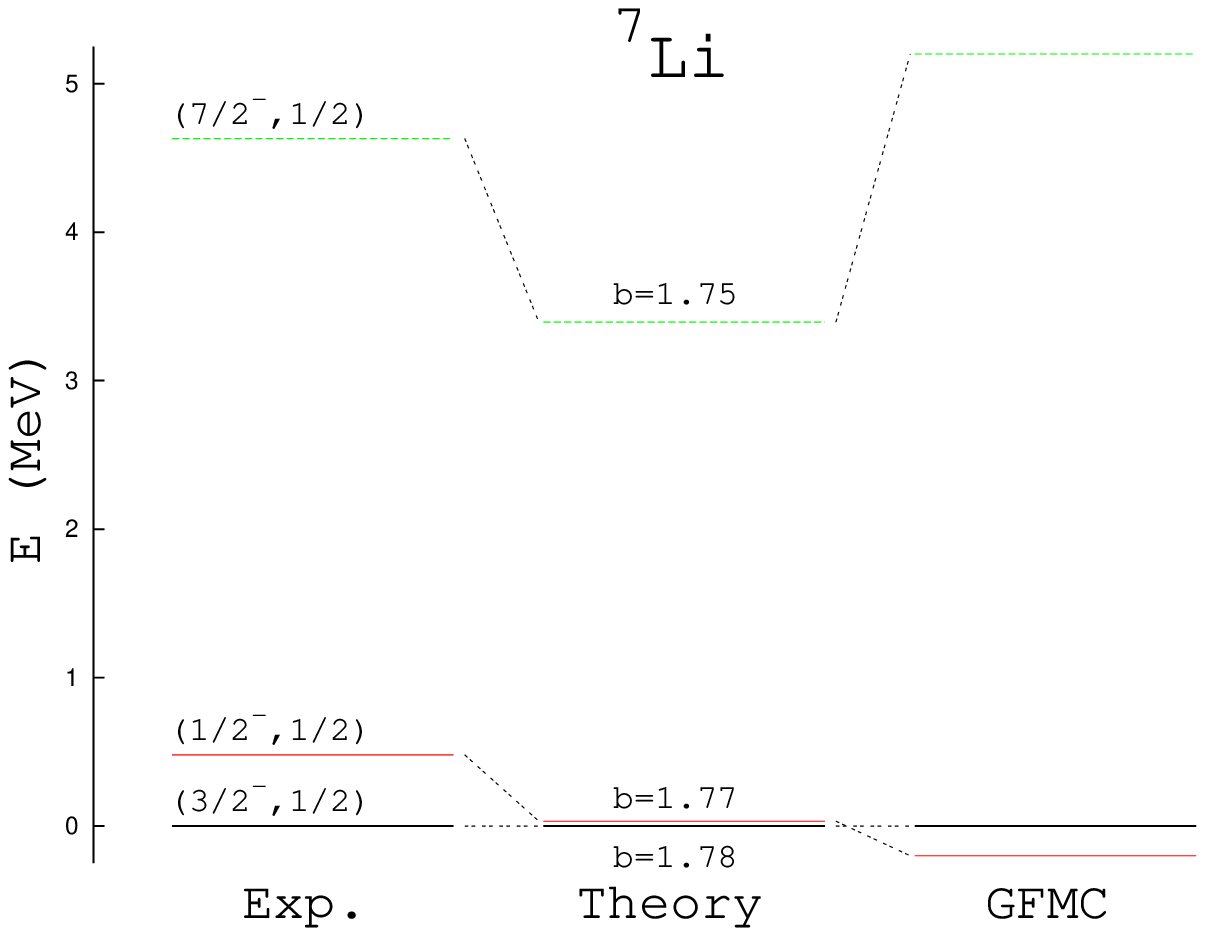,width=12cm}
  \caption{(Color online) Excited spectrum of $^7$Li using
    Av18.\label{fig9}}
\end{figure}

P-shell nuclei have richer structure than their s-shell counterparts
due to the existence of excited states.  In
Figs.~\ref{fig6}-\ref{fig7} we show the excited spectrum of $^6$Li and
$^7$Li using the Av8' potential.  We also show the experimental
spectrum to provide reference points, but stress that our results
should only be compared to GFMC/NCSM results only.  All BH calculations
were done in $\Lambda=0$ included spaces.  The quoted $b$ values are
taken at the minimum binding energy for each level.  The Av18 results
are shown in Figs.~\ref{fig8}-\ref{fig9}.

Our $\Lambda=0$ calculations of excited spectra fall within the range
of GFMC and NCSM results.  We note that the NCSM $^6$Li calculations
were performed in a $\Lambda$=14 included space using $\hbar\Omega=11$
MeV ($b$=1.94 fm) and gave 28.08 MeV binding for the ground state of
$^6$Li using Av8'.  The excitation energies of the $(2^+,0)$ and, in
particular, of the $(1^+_2,0)$ states were not converged yet and are
still decreasing with increasing $\Lambda$.  The NCSM $^7$Li results
are taken from Ref. \cite{Navratil:2003ef}. In general our level
orderings correlate with those of GFMC and NCSM calculations as well.
However, certain levels are inverted, such as the (3$^+$,0)-(0$^+$,1)
states of $^6$Li and the (3/2$^-$,1/2)-(1/2$^-$,1/2) states of $^7$Li.
These particular levels are sensitive to the strength of the
spin-orbit interaction in the $NN$ interaction, which to date, is
poorly represented in all potential models.  Clearly our approximation
of $V_{eff}$ in these limiting included spaces exacerbates this issue.
The energy differences between using Av18 and Av8' are typically less
than $\sim 200$ KeV.

\section{Convergence Issues\label{convergence}}
\begin{figure}
  \epsfig{file=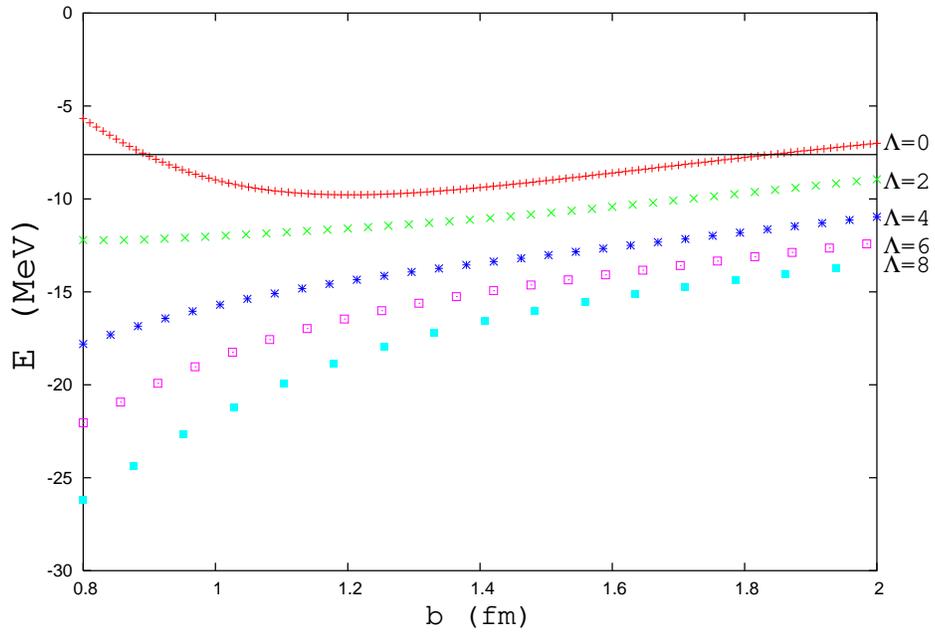,width=12cm}
  \caption{(Color online) Dependence of $^3$H on models space size
  $\Lambda$ and oscillator parameter $b$ using Av18.  The solid line
  is the exact $^3$H binding energy with the Av18
  potential.}\label{fig10}
\end{figure}

The $\Lambda=2$ calculated binding energies for the five- and six-body
systems of Figs.~\ref{fig4}-\ref{fig5} are closer to the exact answers
compared to the $\Lambda=0$ results, and the dependence on $b$ is
somewhat flatter.  This suggests some sort of convergence with
increasing $\Lambda$.  Indeed, as $\Lambda\rightarrow\infty$, one has
that $P\rightarrow 1$, giving $V_{12,eff}\rightarrow V_{12}$.  Hence
in this limit the calculated binding energies are guaranteed to
converge to the exact answers.  However, we caution that since we are
working in such small included spaces, such limiting behavior might
not be seen until $\Lambda$ becomes much larger.  Furthermore, the
argument above does not give any information on how quickly the
binding energies converge to the exact answer.  For example, larger
model space calculations of the $^3$H (Fig.~\ref{fig10}) do not show a
systematic convergence.  This is in direct contrast to NCSM
calculations, where there is definite convergence as the included
space is increased\cite{Navratil:1999pw}.  A possible explanation for
this difference is in the incorporation of a confining HO mean-field
potential in NCSM calculations.  The starting bare Hamiltonian for
NCSM calculations is
\begin{displaymath}
H^{\rm NCSM}=H+T_{c.m.}+U^{\rm HO}_{c.m.},
\end{displaymath}
 where $H$ is given by Eq. (\ref{eqn:h}) and $T_{c.m.}+U^{\rm
HO}_{c.m.}$ is the c.m. harmonic oscillator Hamiltonian.  In
principle, the addition of a confining HO Hamiltonian (which is in the
end subtracted from the effective Hamiltonian in any case) does not
alter the final answer of the many-body problem, as long as the
effective interaction is solved to a high level of accuracy.  Indeed,
we argue that the inclusion of a confining mean-field increases the
rate of convergence as the size of the included-space is increased.
Since our BH calculations lack any confining mean-field (\emph{e.g.}
$U^{\rm HO}_{c.m.}$), our convergence with increasing $\Lambda$ is
hindered.

\section{Conclusion\label{conclusion}}
The exponentially increasing complexity of the many-body Hilbert space
motivates the use of effective interactions in truncated included
spaces.  Unfortunately, the exact effective interaction $H_{eff}$ is
as difficult to deal with as the original infinite-Hilbert space
problem.  Hence some form of approximation must be made to $H_{eff}$,
or more specifically, to $V_{eff}$.  In Sect.~\ref{approx. bh} we
showed a particular approximation of $V_{eff}$ that was obtained by
summing ladder diagrams to all orders, giving $V_{12,eff}$.  We
argued that our approximation of $V_{eff}$, one in which spectator
dependence and cluster recoil are manifestly included, was a natural
extension of the single-valued G-matrix of traditional shell model
calculations and the multi-valued G-matrix of Ref.\cite{Zheng:1995td}.
Our approximation of $V_{eff}$ is equivalent to the leading order term
of the BH expansion of Ref.\cite{Luu:2004xc}.

Because we include the effects of cluster recoil, our approximation of
$V_{eff}$ is an A-body operator, as opposed to the two-body nature of
the G-matrix interactions mentioned above.  In Sect.~\ref{a-body
matrix elements} and Appendix~\ref{int spect} we showed how A-body
matrix elements of $V_{12,eff}$ can be easily calculated by
transforming the A-dimensional nested integrals into a sum over
one-dimensional integrals.  The result is valid for any A-body system
and can be computed quickly and accurately.  We showed that by
integrating over the spectator degrees of freedom, we could define an
average residual interaction for the interacting two-nucleon cluster
that depended on the initial HO momentum configurations of the
spectator nucleons.  Figure~\ref{fig2} shows how this average
interaction in the $^1$S$_0$ channel changes as the number of
spectator nucleons is increased.

Finally, in Sect.~\ref{results} we showed our results of using
$V_{12,eff}$ in s- and some p-shell nuclei for the smallest allowed
included spaces.  All our calculations produced stationary solutions
(\emph{i.e.} negative binding energies).  Our s-shell results show
good agreement with exact theoretical answers.  In particular, the
alpha particle is underbound by only $\sim 1$ MeV.  On the other hand,
our p-shell results all severely underbind in $\Lambda=0$ included
spaces, and are therefore susceptible to cluster breakup.  The
excitation spectra of $^6$Li and $^7$Li fell within the ranges
calculated by GFMC and NCSM methods.  The level orderings correlated
to some extent with these methods as well, though level inversions do
exist.  We argued that these level inversions were closely related to
the weak spin-orbit interactions of the potential models used in our
calculations, which were further exacerbated by our approximation of
$V_{eff}$ in such small included spaces.

For both $^5$He and $^6$Li, we extended our calculations to
$\Lambda=2$ included spaces (see Figs.~\ref{fig3}-\ref{fig4}), and
some semblance of convergence was observed.  We caution, however, that
any apparent convergence in such small spaces is not definitive.
Indeed, for the case of $^3$H, convergence was not seen as the size of
the included-space increased, as shown in Fig.~\ref{fig6}.  This is in
direct contrast to NCSM calculations.  Though our calculations are, in
a sense, more sophisticated than NCSM calculations, our results are no
better for $\Lambda\ge2$.  We argued in Sec.~\ref{convergence} that the
lack of a confining `mean-field' potential was the most likely reason
for our poor convergence.

As mentioned in the previous section, our calculations were not too
computationally demanding since we worked within the smallest allowed
included spaces.  Indeed, we were able to perform all our calculations
on a single Unix box.  Calculations on higher A-body systems, or on
much larger included spaces, will require that our calculations become
parallelized, which we hope to do in the near future.  It would be
interesting to see if the $\Lambda=0$ `degeneracy' persists for larger
p-shell nuclei.  To tackle our convergence issues, we intend to
incorporate a `mean-field' potential into our future calculations.

\appendix
\section{A-body matrix elements of $T_{eff}$\label{Teff abody}}
The relevant expression in Jacobi coordinates is
\begin{multline}\label{eqn:AMea}
  <n'_1l'_1\cdots n'_{A-1}l'_{A-1}|\frac{1}{E-T}|n_1l_1\cdots 
  n_{A-1}l_{A-1}>= 
    \delta_{l'_1,l_1}\cdots\delta_{l'_{A-1},l_{A-1}}\\
    \times\int dp_1\cdots dp_{A-1}
  R^*_{n'_1l_1}(p_1)\cdots R^*_{n'_{A-1}l_{A-1}}(p_{A-1})
  \frac{p_i^2\cdots p_{A-1}^2}{E-\sum_{i=1}^{A-1}\frac{p_i^2}{2M_n}}
  R_{n_1l_1}(p_1)\cdots R_{n_{A-1}l_{A-1}}(p_{A-1}),
\end{multline}
where $R_{nl}(p)$ is the HO radial wavefunction in momentum space.
Replacing the radial integrals by their series expansion and changing
to dimensionless variables $x_1,\ldots,x_{A-1}$, Eq.~\ref{eqn:AMea}
becomes
\begin{multline}\label{eqn:AMeb}
  \delta_{l'_1,l_1}\cdots\delta_{l'_{A-1},l_{A-1}}
  \sum_{m_1,m'_1,\cdots,m_{A-1},m'_{A-1}=0}^{n_1,n'_1,\cdots,n_{A-1},n'_{A-1}}
  A(m_1,m'_1,\cdots,m_{A-1},m'_{A-1})\\ \times \left[\int dx_1\cdots
  dx_{A-1}
  e^{-(x_1^2+\cdots+x_{A-1}^2)}\frac{x_1^{2(l_1+m_1+m_1'+1)}\cdots
  x_{A-1}^{2(l_{A-1}+m_{A-1}+m_{A-1}'+1)}} {E-\frac{1}{2M_n b^2}\sum
  x_i^2}\right],
\end{multline}
where
\begin{multline}
  A(m_1,m'_1,\cdots,m_{A-1},m'_{A-1})=\\
  (-1)^{\Sigma_i(n_i+m_i)}\left[\prod_{i=1}^{A-1}\
    \frac{\sqrt{2\Gamma(n_i+1)\Gamma(n_i+l_i+3/2))}}
    {\Gamma(n_i-m_i+1)\Gamma(l_i+m_i+3/2)\Gamma(m_i+1)}\right]\\
  \times (-1)^{\Sigma_i(n'_i+m'_i)}\left[\prod_{i=1}^{A-1}\
    \frac{\sqrt{2\Gamma(n'_i+1)\Gamma(n'_i+l_i+3/2))}}
    {\Gamma(n'_i-m'_i+1)\Gamma(l_i+m'_i+3/2)\Gamma(m'_i+1)}\right].
\end{multline}
The integral in Eq.~\ref{eqn:AMea} can be evaluated by first changing
variables to (A-1)-dimensional polar coordinates,
$\rho,\theta_{A-5},\ldots,\theta_1,\phi$, where $0\le\phi\le 2\pi$,
and $0\le\theta_i\le\pi$:
\begin{equation}\label{eqn:AMec}
  \begin{aligned}
    x_1=&\ \rho\ cos\phi\ sin\theta_1\ sin\theta_2\cdots sin\theta_{A-5}\\
    x_2=&\ \rho\ sin\phi\ sin\theta_1\ sin\theta_2\cdots sin\theta_{A-5}\\
    x_3=&\ \rho\cos\theta_1\ sin\theta_2\cdots sin\theta_{A-5}\\
    \vdots=&\ \vdots\\
    x_i=&\ \rho\ cos\theta_{i-2}\ sin\theta_{i-1}\cdots sin\theta_{A-5}\\
    \vdots=&\ \vdots\\
    x_{A-1}=&\ \rho\ cos\theta_{A-5}\\
    \rho^2=&\ \sum x_i^2.
    \end{aligned}
\end{equation}
The infinitesimal volume element is
\begin{displaymath}
  dx_1\cdots dx_{A-1}=\ \rho^{A-4}\ sin^{A-5}\theta_{A-5}\cdots
  sin\theta_1\ d\rho\ d\theta_{A-5}\cdots d\theta_1\ d\phi.
\end{displaymath}
With these change of variables, the integral in
Eq.~\ref{eqn:AMea} becomes
\begin{multline}\label{eqn:AMed}
  \left[\int_0^{\pi/2}(cos\phi)^{2(l_1+m_1+m_1'+1)}
    (sin\phi)^{2(l_2+m_2+m_2'+1)}d\phi\right]\\
  \times\left[\int_0^{\pi/2}(cos\theta_1)^{2(l_3+m_3+m_3'+1)}
    (sin\theta_1)^{2\sum_{i=1}^2(l_i+m_i+m_i'+1)+1}d\theta_1\right]\\
  \cdots\times\left[\int_0^{\pi/2}(cos\theta_{A-5})^{2(l_{A-1}+m_{A-1}+m_{A-1}'
      +1)}
    (sin\theta_{A-5})^{2\sum_{i=1}^{A-2}(l_i+m_i+m_i'+1)+A-5}d
    \theta_{A-5}\right]\\
  \times\left[\int_0^{\infty} d\rho\
    e^{-\rho^2}\frac{\rho^{A-2+2\sum_{i=1}^{A-1}(l_i+m_i+m_i'+1)}}
    {E-\frac{\rho^2}{2M_n b^2}}\right].
\end{multline}
Note the limits of integration.  All the angular integrals and the
integral over $\rho$ can be expressed analytically,
\begin{eqnarray}
  \int_o^{\pi/2}d\theta\ cos^{\alpha}\theta\ sin^{\beta}\theta&=&
  \frac{\Gamma(\frac{\alpha+1}{2})\Gamma(\frac{\beta+1}{2})}
  {2\Gamma(\frac{\alpha+\beta}{2}+1)},\\\label{eqn:angular}
  \int_0^\infty d\rho\
  e^{-\rho^2}\frac{\rho^{2\alpha+1}}{\epsilon-\rho^2}&=&
  \frac{(-1)^{\alpha+1}}{2}\epsilon^\alpha e^{-\epsilon}\ 
  \Gamma(-\alpha,-\epsilon)
  \Gamma(\alpha+1).\label{eqn:rho}
\end{eqnarray}
Plugging Eqs.~\ref{eqn:AMed}-~\ref{eqn:rho} into Eq.~\ref{eqn:AMeb}
gives the final result,
\begin{multline}\label{AMee}
  <n'_1l'_1\cdots n'_{A-1}l'_{A-1}|\frac{1}{E-T}|n_1l_1\cdots 
  n_{A-1}l_{A-1}>=\\
  \delta_{l'_1,l_1}\cdots\delta_{l'_{A-1},l_{A-1}}
  \frac{2}{\hbar\omega}
  \sum_{m_1,m'_1,\cdots,m_{A-1},m'_{A-1}=0}^{n_1,n'_1,\cdots,n_{A-1},n'_{A-1}}
  B(m_1,m'_1,\cdots,m_{A-1},m'_{A-1})\\
  \times\left[\frac{(-1)^{\alpha+1}}{2}
    \left(\frac{2E}{\hbar\omega}\right)^\alpha 
  e^{-\frac{2E}{\hbar\omega}}\ 
  \Gamma(-\alpha,-\frac{2E}{\hbar\omega})\Gamma(\alpha+1)\right],
\end{multline}
where $\alpha=(A-3)/2+\sum_i(l_i+m_i+m'_i+1)$ and
\begin{multline}\label{eqn:Bcoeff}
  B(m_1,m'_1,\cdots,m_{A-1},m'_{A-1})=\\
  (-1)^{\Sigma_i(n_i+m_i)}\left[\prod_{i=1}^{A-1}\
    \frac{\sqrt{2\Gamma(n_i+1)\Gamma(n_i+l_i+3/2))}}
    {\Gamma(n_i-m_i+1)\Gamma(l_i+m_i+3/2)\Gamma(m_i+1)}\right]\\
  \times (-1)^{\Sigma_i(n'_i+m'_i)}\left[\prod_{i=1}^{A-1}\
    \frac{\sqrt{2\Gamma(n'_i+1)\Gamma(n'_i+l_i+3/2))}}
    {\Gamma(n'_i-m'_i+1)\Gamma(l_i+m'_i+3/2)\Gamma(m'_i+1)}\right]\\
  \times\left[  \prod_{i=2}^{A-1}
    \frac{\Gamma(l_i+m_i+m_i'+3/2)\Gamma(\sum_{j=1}^{i-1}(l_j+m_j+m_j'+1)+
      \frac{i-1}{2})}
    {2\Gamma(l_i+m_i+m_i'+\sum_{j=1}^{i-1}(l_j+m_j+m_j'+1)+\frac{i+2}{2})}
    \right].
\end{multline}

\section{Integrating out spectator degrees of freedom\label{int spect}}
The relevant expression here is 
\begin{equation}\label{eqn:BMea}
  <p';n_1'l_1',\cdots,n_{A-2}'l_{A-2}'|\hat{F}
  |p;n_1l_1,\cdots,n_{A-2}l_{A-2}>,
\end{equation}
where the interacting two-nucleon initial and final momenta are $p$
and $p'$, respectively, and the spectator quantum numbers are denoted
with subscripts that run from $1$ through $A-2$..  $\hat{F}$ is some
operator that depends only on $p$, $p'$, and $\sum_i q_i^2$, where
$q_i$ are the spectator momenta.  $G_0$, $t_{12}$, and $G_0t_{12}G_0$
are three examples of such an operator.  Since $F$ only depends on
$q_i^2$, which is a scalar, it cannot change the orbital quantum
numbers of the spectator nucleons.  Hence Eq.~\ref{eqn:BMea} becomes
\begin{multline}\label{eqn:BMeb}
  \delta_{l_1',l_1}\cdots\delta_{l_{A-2}',l_{A-2}} \\
  \times\int dq_1\cdots dq_{A-2} q_1^2\cdots q_{A-2}^2
  R^*_{n_1'l_1}(q_1)R_{n_1l_1}(q_1)\cdots
  R^*_{n_{A-2}'l_{A-2}}(q_{A-2})R_{n_{A-2}l_{A-2}}(q_{A-2})\\
  \times F(p,p';\sum_{i=1}^{A-2}q_i^2).
\end{multline}
Equation~\ref{eqn:BMeb} is very similar to Eq.~\ref{eqn:AMea}.
Indeed, we can follow a similar procedure outlined in the previous
section to come to our desired result.  Repeating the steps from
Eq.~\ref{eqn:AMeb} through Eq.~\ref{eqn:angular}, but with a change of
variables to (A-2)-dimensional polar coordinates rather than (A-1),
gives 
\begin{multline}\label{eqn:BMec}
  \delta_{l_1',l_1}\cdots\delta_{l_{A-2}',l_{A-2}}
  \sum_{m_1,m'_1,\cdots,m_{A-2},m'_{A-2}=0}^{n_1,n'_1,\cdots,n_{A-2},n'_{A-2}}
  C(m_1,m'_1,\cdots,m_{A-2},m'_{A-2})\\
  \times\left[\int_0^{\infty} d\rho\ \rho^{A-3+2\Sigma_i^{A-2} 
      (l_i+m_i+m_i'+1)}e^{-\rho^2}
    F(p',p;\frac{\rho^2}{b^2})\right],
\end{multline}
where $b$ is the oscillator parameter and
\begin{multline}\label{eqn:CCcoeff}
  C(m_1,m'_1,\cdots,m_{A-2},m'_{A-2})=\\
  (-1)^{\Sigma_i(n_i+m_i)}\left[\prod_{i=1}^{A-2}\
    \frac{\sqrt{2\Gamma(n_i+1)\Gamma(n_i+l_i+3/2))}}
    {\Gamma(n_i-m_i+1)\Gamma(l_i+m_i+3/2)\Gamma(m_i+1)}\right]\\
  \times (-1)^{\Sigma_i(n'_i+m'_i)}\left[\prod_{i=1}^{A-2}\
    \frac{\sqrt{2\Gamma(n'_i+1)\Gamma(n'_i+l_i+3/2))}}
    {\Gamma(n'_i-m'_i+1)\Gamma(l_i+m'_i+3/2)\Gamma(m'_i+1)}\right]\\
  \times\left[  \prod_{i=2}^{A-2}
    \frac{\Gamma(l_i+m_i+m_i'+3/2)\Gamma(\sum_{j=1}^{i-1}(l_j+m_j+m_j'+1)+
      \frac{i-1}{2})}
    {2\Gamma(l_i+m_i+m_i'+\sum_{j=1}^{i-1}(l_j+m_j+m_j'+1)+\frac{i+2}{2})}
    \right].
\end{multline}
Note that the integral in Eq.~\ref{eqn:BMec} cannot be determined
analytically, as in the previous section.  However, it can be
numerically evaluated quickly and accurately.

\begin{acknowledgments}
  This work was performed in part under the auspices of the
  U. S. Department of Energy by the University of California, Lawrence
  Livermore National Laboratory under contract No. W-7405-Eng-48, and
  in part under DOE grants No. DE-FC02-01ER41187 and
  DE-FG02-00ER41132.
\end{acknowledgments}

\newpage
%Create the reference section using BibTeX:
%\bibliography{references}

\begin{thebibliography}{17}
\expandafter\ifx\csname natexlab\endcsname\relax\def\natexlab#1{#1}\fi
\expandafter\ifx\csname bibnamefont\endcsname\relax
  \def\bibnamefont#1{#1}\fi
\expandafter\ifx\csname bibfnamefont\endcsname\relax
  \def\bibfnamefont#1{#1}\fi
\expandafter\ifx\csname citenamefont\endcsname\relax
  \def\citenamefont#1{#1}\fi
\expandafter\ifx\csname url\endcsname\relax
  \def\url#1{\texttt{#1}}\fi
\expandafter\ifx\csname urlprefix\endcsname\relax\def\urlprefix{URL }\fi
\providecommand{\bibinfo}[2]{#2}
\providecommand{\eprint}[2][]{\url{#2}}

\bibitem[{\citenamefont{Pieper and Wiringa}(2001)}]{Pieper:2001mp}
\bibinfo{author}{\bibfnamefont{S.~C.} \bibnamefont{Pieper}} \bibnamefont{and}
  \bibinfo{author}{\bibfnamefont{R.~B.} \bibnamefont{Wiringa}},
  \bibinfo{journal}{Ann. Rev. Nucl. Part. Sci.} \textbf{\bibinfo{volume}{51}},
  \bibinfo{pages}{53} (\bibinfo{year}{2001}), \eprint{nucl-th/0103005}.

\bibitem[{\citenamefont{Carlson and Schiavilla}(1998)}]{Carlson:1998qn}
\bibinfo{author}{\bibfnamefont{J.}~\bibnamefont{Carlson}} \bibnamefont{and}
  \bibinfo{author}{\bibfnamefont{R.}~\bibnamefont{Schiavilla}},
  \bibinfo{journal}{Rev. Mod. Phys.} \textbf{\bibinfo{volume}{70}},
  \bibinfo{pages}{743} (\bibinfo{year}{1998}).

\bibitem[{\citenamefont{Pudliner et~al.}(1997)\citenamefont{Pudliner,
  Pandharipande, Carlson, Pieper, and Wiringa}}]{Pudliner:1997ck}
\bibinfo{author}{\bibfnamefont{B.~S.} \bibnamefont{Pudliner}},
  \bibinfo{author}{\bibfnamefont{V.~R.} \bibnamefont{Pandharipande}},
  \bibinfo{author}{\bibfnamefont{J.}~\bibnamefont{Carlson}},
  \bibinfo{author}{\bibfnamefont{S.~C.} \bibnamefont{Pieper}},
  \bibnamefont{and} \bibinfo{author}{\bibfnamefont{R.~B.}
  \bibnamefont{Wiringa}}, \bibinfo{journal}{Phys. Rev.}
  \textbf{\bibinfo{volume}{C56}}, \bibinfo{pages}{1720} (\bibinfo{year}{1997}),
  \eprint{nucl-th/9705009}.

\bibitem[{\citenamefont{Bruckner}(1995{\natexlab{a}})}]{Bruckner:1955a}
\bibinfo{author}{\bibfnamefont{K.}~\bibnamefont{Bruckner}},
  \bibinfo{journal}{Phys. Rev.} \textbf{\bibinfo{volume}{97}},
  \bibinfo{pages}{1353} (\bibinfo{year}{1995}{\natexlab{a}}).

\bibitem[{\citenamefont{Bruckner}(1995{\natexlab{b}})}]{Bruckner:1955b}
\bibinfo{author}{\bibfnamefont{K.}~\bibnamefont{Bruckner}},
  \bibinfo{journal}{Phys. Rev.} \textbf{\bibinfo{volume}{100}},
  \bibinfo{pages}{36} (\bibinfo{year}{1995}{\natexlab{b}}).

\bibitem[{\citenamefont{Navratil et~al.}(2000)\citenamefont{Navratil,
  Kamuntavicius, and Barrett}}]{Navratil:1999pw}
\bibinfo{author}{\bibfnamefont{P.}~\bibnamefont{Navratil}},
  \bibinfo{author}{\bibfnamefont{G.~P.} \bibnamefont{Kamuntavicius}},
  \bibnamefont{and} \bibinfo{author}{\bibfnamefont{B.~R.}
  \bibnamefont{Barrett}}, \bibinfo{journal}{Phys. Rev.}
  \textbf{\bibinfo{volume}{C61}}, \bibinfo{pages}{044001}
  (\bibinfo{year}{2000}), \eprint{nucl-th/9907054}.

\bibitem[{\citenamefont{Navratil and Ormand}(2003)}]{Navratil:2003ef}
\bibinfo{author}{\bibfnamefont{P.}~\bibnamefont{Navratil}} \bibnamefont{and}
  \bibinfo{author}{\bibfnamefont{W.~E.} \bibnamefont{Ormand}},
  \bibinfo{journal}{Phys. Rev.} \textbf{\bibinfo{volume}{C68}},
  \bibinfo{pages}{034305} (\bibinfo{year}{2003}), \eprint{nucl-th/0305090}.

\bibitem[{\citenamefont{Navratil and Caurier}(2003)}]{Navratil:2003ib}
\bibinfo{author}{\bibfnamefont{P.}~\bibnamefont{Navratil}} \bibnamefont{and}
  \bibinfo{author}{\bibfnamefont{E.}~\bibnamefont{Caurier}}
  (\bibinfo{year}{2003}), \eprint{nucl-th/0311036}.

\bibitem[{\citenamefont{Bloch and Horowitz}(1958)}]{bloch:1958}
\bibinfo{author}{\bibfnamefont{C.}~\bibnamefont{Bloch}} \bibnamefont{and}
  \bibinfo{author}{\bibfnamefont{J.}~\bibnamefont{Horowitz}},
  \bibinfo{journal}{Nucl. Phys.} \textbf{\bibinfo{volume}{8}},
  \bibinfo{pages}{91} (\bibinfo{year}{1958}).

\bibitem[{\citenamefont{Luu et~al.}(2004)\citenamefont{Luu, Bogner, Haxton, and
  Navratil}}]{Luu:2004xc}
\bibinfo{author}{\bibfnamefont{T.~C.} \bibnamefont{Luu}},
  \bibinfo{author}{\bibfnamefont{S.}~\bibnamefont{Bogner}},
  \bibinfo{author}{\bibfnamefont{W.~C.} \bibnamefont{Haxton}},
  \bibnamefont{and} \bibinfo{author}{\bibfnamefont{P.}~\bibnamefont{Navratil}},
  \bibinfo{journal}{Phys. Rev.} \textbf{\bibinfo{volume}{C70}},
  \bibinfo{pages}{014316} (\bibinfo{year}{2004}), \eprint{nucl-th/0404028}.

\bibitem[{\citenamefont{Zheng et~al.}(1995)\citenamefont{Zheng, Barrett, Vary,
  Haxton, and Song}}]{Zheng:1995td}
\bibinfo{author}{\bibfnamefont{D.~C.} \bibnamefont{Zheng}},
  \bibinfo{author}{\bibfnamefont{B.~R.} \bibnamefont{Barrett}},
  \bibinfo{author}{\bibfnamefont{J.~P.} \bibnamefont{Vary}},
  \bibinfo{author}{\bibfnamefont{W.~C.} \bibnamefont{Haxton}},
  \bibnamefont{and} \bibinfo{author}{\bibfnamefont{C.~L.} \bibnamefont{Song}},
  \bibinfo{journal}{Phys. Rev.} \textbf{\bibinfo{volume}{C52}},
  \bibinfo{pages}{2488} (\bibinfo{year}{1995}).

\bibitem[{\citenamefont{Haxton and Luu}(2002)}]{Haxton:2002kb}
\bibinfo{author}{\bibfnamefont{W.~C.} \bibnamefont{Haxton}} \bibnamefont{and}
  \bibinfo{author}{\bibfnamefont{T.}~\bibnamefont{Luu}},
  \bibinfo{journal}{Phys. Rev. Lett.} \textbf{\bibinfo{volume}{89}},
  \bibinfo{pages}{182503} (\bibinfo{year}{2002}), \eprint{nucl-th/0204072}.

\bibitem[{\citenamefont{Fetter and Walecka}(2003)}]{Fetter&Walecka:1971}
\bibinfo{author}{\bibfnamefont{A.}~\bibnamefont{Fetter}} \bibnamefont{and}
  \bibinfo{author}{\bibfnamefont{J.}~\bibnamefont{Walecka}},
  \emph{\bibinfo{title}{Quantum Theory of Many-Particle Systems}}
  (\bibinfo{publisher}{Dover Publications}, \bibinfo{address}{New York},
  \bibinfo{year}{2003}).

\bibitem[{\citenamefont{Barrett and Kirson}(1970)}]{Barrett:1970}
\bibinfo{author}{\bibfnamefont{B.}~\bibnamefont{Barrett}} \bibnamefont{and}
  \bibinfo{author}{\bibfnamefont{M.}~\bibnamefont{Kirson}},
  \bibinfo{journal}{Nucl.Phys.} \textbf{\bibinfo{volume}{A178}},
  \bibinfo{pages}{145} (\bibinfo{year}{1970}).

\bibitem[{\citenamefont{Wiringa et~al.}(1995)\citenamefont{Wiringa, Stoks, and
  Schiavilla}}]{Wiringa:1995wb}
\bibinfo{author}{\bibfnamefont{R.~B.} \bibnamefont{Wiringa}},
  \bibinfo{author}{\bibfnamefont{V.~G.~J.} \bibnamefont{Stoks}},
  \bibnamefont{and}
  \bibinfo{author}{\bibfnamefont{R.}~\bibnamefont{Schiavilla}},
  \bibinfo{journal}{Phys. Rev.} \textbf{\bibinfo{volume}{C51}},
  \bibinfo{pages}{38} (\bibinfo{year}{1995}), \eprint{nucl-th/9408016}.

\bibitem[{\citenamefont{Nogga et~al.}(2002)\citenamefont{Nogga, Kamada,
  Gloeckle, and Barrett}}]{Nogga:2001cz}
\bibinfo{author}{\bibfnamefont{A.}~\bibnamefont{Nogga}},
  \bibinfo{author}{\bibfnamefont{H.}~\bibnamefont{Kamada}},
  \bibinfo{author}{\bibfnamefont{W.}~\bibnamefont{Gloeckle}}, \bibnamefont{and}
  \bibinfo{author}{\bibfnamefont{B.~R.} \bibnamefont{Barrett}},
  \bibinfo{journal}{Phys. Rev.} \textbf{\bibinfo{volume}{C65}},
  \bibinfo{pages}{054003} (\bibinfo{year}{2002}), \eprint{nucl-th/0112026}.

\bibitem[{\citenamefont{Pieper et~al.}(2004)\citenamefont{Pieper, Wiringa, and
  Carlson}}]{Pieper:2004qw}
\bibinfo{author}{\bibfnamefont{S.~C.} \bibnamefont{Pieper}},
  \bibinfo{author}{\bibfnamefont{R.~B.} \bibnamefont{Wiringa}},
  \bibnamefont{and} \bibinfo{author}{\bibfnamefont{J.}~\bibnamefont{Carlson}},
  \bibinfo{journal}{Phys. Rev.} \textbf{\bibinfo{volume}{C70}},
  \bibinfo{pages}{054325} (\bibinfo{year}{2004}), \eprint{nucl-th/0409012}.

\end{thebibliography}

\newpage

%% Tables below. . .
\begin{table}[h]
\caption{Dimensions of $\Lambda=0,2$ included spaces for different 
  few-body ground states.\label{tab:dimensions}}
\begin{ruledtabular}
\begin{tabular}{c c c c c c c}
$\Lambda$ & $^2$H & $^3$H/$^3$He & $^4$He & $^5$He & $^6$Li & $^7$Li
 \\ \hline 0 & 1 & 1 & 1 & 1 & 3 & 5\\ 2 & 3 & 5 & 5 & 26 & 48 & 185 \\
\end{tabular}
\end{ruledtabular}
\end{table}

% Figures below. . .

\end{document}